\documentclass{Interspeech2024}
\usepackage{bbold}

\interspeechcameraready

\title{Less Forgetting for Better Generalization: Exploring Continual-learning Fine-tuning Methods for Speech Self-supervised Representations}

\name[affiliation={1}]{Salah}{Zaiem}
\name[affiliation={2}]{Titouan}{Parcollet}
\name[affiliation={1}]{Slim}{Essid}

\address{
  $^1$LTCI, Télécom Paris, Institut Polytechnique de Paris, France\\
  $^2$Samsung AI Research, Cambridge, United-Kingdom}
\email{zaiemsalah@gmail.com, parcollet@samsung.com, slim.essid@telecom-paris.fr}


\begin{document}

\maketitle

\begin{abstract}
    
Despite being trained on massive and diverse datasets, speech self-supervised encoders are generally used for downstream purposes as mere frozen feature extractors or model initializers before fine-tuning. The former severely limits the exploitation of large encoders, while the latter hurts the robustness acquired during pretraining, especially in low-resource scenarios. This work explores middle-ground solutions, conjecturing that reducing the forgetting of the self-supervised task during the downstream fine-tuning leads to better generalization. To prove this, focusing on speech recognition, we benchmark different continual-learning approaches during fine-tuning and show that they improve both in-domain and out-of-domain generalization abilities. Relative performance gains reach $15.7\%$ and $22.5\%$ with XLSR used as the encoder on two English and Danish speech recognition tasks.  Further probing experiments show that these gains are indeed linked to less forgetting.
\end{abstract}

\section{Introduction}
Self-supervised representations are often used in low-resource scenarios where downstream data can consist of only a handful of hours of annotated audio \cite{shi23g_interspeech, 9801640}. During evaluation and benchmarking, self-supervised encoders are generally kept frozen\cite{zaiem23b_interspeech, superb}. However, in common uses, the encoder weights are also fine-tuned. On the one hand, freezing the self-supervised representations during downstream training makes the SSL backbone a mere feature extractor. In this case, to reach reasonable performance, the downstream head may need to be more complex leading to costly inferences \cite{zaiem2023speech}. On the other hand, full fine-tuning of the SSL encoder reduces the pretraining to a superior network initialization.

In this work, we postulate that fine-tuning the whole network weights hurts the generalization abilities of the final obtained model, because the model may ``forget" what has been learned in the pretraining phase. While this has been proven in text-related applications \cite{recadam,he-etal-2021-analyzing}, it has not been explicitly studied in previous speech-processing works. The main argument for the postulate is that models generally learn to solve the self-supervision on massive unlabeled datasets. This large data diversity makes these models robust to distributional shifts and explains in large part their generalization abilities across different domains \cite{hendrycks2019using} and should thus be kept after the fine-tuning.

Forgetting issues have been tackled in the Continual Learning (CL) literature. CL is a machine learning paradigm that focuses on training models to acquire new knowledge, adapting to new tasks and data, without losing prior abilities \cite{PARISI201954}. This work explores the use of continual-learning methods during downstream training, keeping, even after fine-tuning, pretraining task abilities. 
Continual learning approaches have been explored lately for speech processing towards including, within a given model scope, new languages \cite{dellalibera2023clmasr, hou2021exploiting}, new accents \cite{trinh22_interspeech,majumdar} or new speakers \cite{Diwan_2023} without losing previous abilities. For self-supervision purposes, it has been used to further train an SSL model to include new domains where the learned representations can be efficiently used for downstream training \cite{lee22i_interspeech}. However, these works never explored CL usage during fine-tuning.

A close line of work is parameter-efficient fine-tuning (PEFT) \cite{10096311}. While reducing the number of parameters updated is done mainly for the sake of efficiency in the case of large pretrained models, it also leads to less forgetting through freezing large parts of the network \cite{ren2024analyzing}. Those methods are widely adopted in the natural language processing and computer vision communities due to the large size of the models \cite{hu2022lora}. The PEFT speech-related literature is more scarce even if positive results were obtained for child-directed speech \cite{9864219} or emotion recognition \cite{peftser}. One close effort has tried various adapter-based methods for self-supervised models on a group of speech tasks \cite{10023274}. However, it focuses on training efficiency, while we show its impact on performance, especially on out-of-distribtion (OOD) testing samples. 

Finally, it is important to note that in our case, while we use methods inspired by continual learning, downstream task performance, in and out-of-domain, is the only objective. We do not evaluate the methods proposed on their forgetting reduction capacity as in the classic CL literature. The link between forgetting and performance is only probed in a second time. The contributions of this work are, thus, three-fold: 

\begin{enumerate}
    \item We explore several continual-learning-based approaches for speech SSL fine-tuning showing substantial performance both on in-domain and out-of-domain testing samples. In particular, as far as we know, we are the first to explore replay options in this context. (Sections \ref{sec:methods} and \ref{sec:exps}).
    \item We highlight the link between the performance gain and the non-forgetting of the self-supervised task by probing the forgetting of the best-performing methods. (Section \ref{sec:analysis}).
    \item We release the SpeechBrain-based \cite{speechbrain} \footnote{github.com/salah-zaiem/CL\_based\_FT}code to enable reproduction of our work and further investigations. 
\end{enumerate}

\section{Methods} \label{sec:methods}
This section introduces the downstream fine-tuning methods we compare in this work. After describing a few classic baseline approaches, it moves to continual-learning-inspired ones, divided into two groups: freezing and replay-based approaches. 

\subsection{Baselines} 
Four baselines, inspired by common practice, are considered in this work. The first one is the vanilla full fine-tuning of the network with the downstream task loss. For the second baseline, the convolutional front-end, often called "acoustic feature extractor" is frozen during fine-tuning. It is more common than full fine-tuning in the speech self-supervision literature as masking in the self-supervised pretraining happens generally after the convolutional front-end \cite{10193042}.

The third baseline targets a better initialization of the downstream head by keeping the encoder frozen during the first steps of downstream training \cite{kumar2022fine}. In a second time, after the initialization of the downstream head, the weights of the encoder are unfrozen, and the whole model is fine-tuned. This method will be called ``two-phased" in the following and has been shown, in computer vision, to lead to better generalization \cite{kumar2022fine}. The final baseline consists in freezing the SSL encoder during the whole downstream finetuning. In this case and following common practices in frozen SSL benchmarking \cite{superb, zaiem2023speech}, the layer input to the downstream head is a weighted sum of the encoder layers, with weights learned during fine-tuning.

\subsection{Freezing-Based Approaches} This section presents a group of tested fine-tuning methods grouped under the ``freezing-based" title as they consist in freezing totally or partially a group of weights learned in the pretraining phase. Among these, we present and test three methods.

\noindent \textbf{Adapters.} The first one uses adapters within the encoder layers \cite{majumdar}. Adapters are lightweight modules intervening after the dense layers that come after self-attention. Precisely, instead of feeding to the next encoder layer the output of the feed-forward layer following the attention, this output is passed through the adapter and summed to itself as in residual approaches. Only the weights of the adapters are changed during the fine-tuning.

\noindent \textbf{LoRa.} Second, following successful trends in natural language processing and computer vision, we test Low-Rank \cite{hu2022lora} (LoRa) fine-tuning. It consists of freezing the pretrained model weights and injecting trainable rank decomposition matrices into each layer of the Transformer architecture, reducing the number of trainable parameters for downstream tasks. Precisely, we replace the feed-forward layers after the self-attention mechanism with LoRA layers. The initial matrix $W_0 \in \mathcal{R}^{d\times k}$  is replaced with a low-rank decomposition $W_0 + \Delta W $ with $\Delta W = BA$ where $B \in \mathcal{R}^{d \times r}$ and $A \in \mathcal{R}^{r \times k}$ with $r$ the rank of the low-rank matrix and $d,k$ the dimensions of latent representations.

\noindent \textbf{Elastic Weight Consolidation.} Finally, we explore using Elastic Weight Consolidation (EWC) \cite{kirkpatrick2017overcoming} during fine-tuning. EWC fine-tuning implies an additional loss, during downstream training, that forces the weights of the final model to be closer to those at the end of the pretraining phase. For every updated parameter, the distance to the initial phase is penalized by the corresponding Fisher information value. The loss becomes: 
\begin{equation} \label{eq:ewc}
    \mathcal{L}(\theta)=\mathcal{L}_{DS}(\theta)+\sum_i \frac{\lambda}{2} F_i\left(\theta_i-\theta_{i}^*\right)^2
\end{equation}

with $\theta$ the parameters of the SSL model, $ \mathcal{L}_{DS}$ the downstream loss, $\theta^*$ the frozen SSL model weights after pretraining, $F$ the Fisher information matrix and $\lambda$ a weighting hyper-parameter. The Fisher information matrix \cite{Fisher22:MathFound} captures how sensible is the pretraining loss to a given parameter. Thus, the loss above penalizes the movement of the most important parameters to the self-supervision task, leading to less forgetting.

\subsection{Replay-Based Approaches}  We also explore replay methods, often called ``experience replay" \cite{rolnick2019experience} in the continual learning literature, during the fine-tuning of self-supervised representations. Replaying the pretraining task explicitly enforces non-forgetting through optimizing for simultaneously low SSL and downstream losses. We noticed that the replay loss should have a lower weight than the downstream one for optimal performance. So, for faster training, instead of weighting the replay loss, at every training step, we sample a random variable $Y \sim U(0, 1)$ and load a replay batch with a probability $p_R$ to perform the replay task. The fine-tuning loss becomes : 
\begin{equation} \label{eq:replay}
        \mathcal{L}(\theta)=\mathcal{L}_{DS, X_{DS}}(\theta)+\mathbb{1}(Y < p_R)  \mathcal{L}_{SSL, X_{R}}(\theta)
\end{equation}
with $\mathcal{L}_{DS}$ the downstream ASR loss,  $\mathcal{L}_{SSL}$ the self-supervision loss and $p_R$ the replay frequency.  Every loss is associated in the formula to the source dataset of the batch it is performed on, with $X_{DS}$ the downstream dataset and $X_R$ the replay one.

\begin{table*}[]
\centering
\scalebox{0.70}{\begin{tabular}{c|lccccccc|ccc} \toprule
&\textbf{Method}         & \multicolumn{7}{c}{\textbf{English Training}}       &\multicolumn{3}{c}{\textbf{Danish Training}}        \\ \toprule

      &   & GS Test        & LS test-clean & LS test-other  & Scottish       & Welsh          & CV &Mean English  & NST Test& CV  &Mean Danish         \\ \toprule
 \textbf{Data2Vec Base}& 
 Baselines      &                &               &                &                &                &                \\ \midrule
&Frozen         & 33.38          & 17            & 22.81          & 38.05          & 33.22          & 56.12&33.43    &70.35 & 83.57 & 76.96    \\
&Full FT        & 26.92          & 9.83          & 17.47          & 26.9           & 22.32          & 53.4 &26.14&13.75&36.57&   25.16       \\
&Fixed CNN      & 26.67          & 10.01         & 16.94          & 25.52          & 22.65          & 49.98  &25.30  &13.8&34.38 &24.09     \\
&Two-Phase      & 26.67          & 10.14         & 17.71          & 26.28          & 23.65          & 49.1  &25.59 &14.63&36.56 &25.60        \\\cmidrule{2-12}
&Freezing-Based &                &               &                &                &                &                \\ \cmidrule{2-12}
&LoRa           & 25.74          & \textbf{9.27} & \textbf{15.73} & 25.18          & 21.88          & 50.81 &24.76&\textbf{12.89}& \textbf{31.13}    &\textbf{22.01}     \\
&EWC            & \textbf{25.57} & 9.4           & 16.3           & \textbf{24.97} & 21.08          & 50.11 &24.57 &12.95&31.70 &22.33       \\
&Adapters       & 30.62          & 12.81         & 19.72          & 35.16          & 30.8           & 56.42  & 30.92& 45.48 & 62.43 &53.96     \\
 \cmidrule{2-12}
&Replay         &                &               &                &                &                &                \\ \cmidrule{2-12}
&LS-Replay   & 26.07          & 9.71          & 16.34          & 25.14          & \textbf{20.37} & \textbf{48.35} & \textbf{24.33}& 12.93 & 32.36& 22.64\\
&Auto-Replay    & 26.25	&9.54&	17.16&	25.8	&22.91&	50.48 &25.36& 13.14 & 35.93 &   24.54     \\ \midrule

\textbf{XLSR-53}&Baselines      &                &               &                &                &                &                \\ \midrule
&Frozen         & $>$ 100          & $>$100            & $>$100          & $>$100          & $>$100          &$>$100& N/A    &$>$100 & $>$100   &N/A   \\
&Full FT        & 28.85          & 11.89          &24.43&32.35&28.42&60.69&31.10  &10.99&30.41 &20.07         \\
&Fixed CNN      &28.98&12&24.35&33.49&29.42&58.88    &31.10&10.8&27.87& 19.34    \\
&Two-Phase      &27.42&10.97&21.66&30.23&25.08&56.19   &28.59&11.21&28.94 & 20.07   \\\cmidrule{2-12}
&Freezing-Based &                &               &                &                &                &                \\ \cmidrule{2-12}
&LoRa           &\textbf{26.68}&10.73&\textbf{19.79}&\textbf{28.61}&\textbf{24.02}&50.83 &\textbf{26.78}&10.37&24.7  &   17.54    \\
&EWC            & 27.21&\textbf{10.55}&20.14&29.58&27.02& 51.12  & 27.60& 10.35&24.44 &17.40     \\
&Adapters       &28.8&12.76&20.3&29.05&26.36&\textbf{50.61}&27.98& 18.85&33.34  & 26.10  \\
 \cmidrule{2-12}
&Replay         &                &               &                &                &                &                \\ \cmidrule{2-12}
&LS-Replay   & 27.54&10.85&20.21&29.15&27.53&53.98 & 28.21&\textbf{9.29}& \textbf{23.56} & \textbf{16.43}\\
&Auto-Replay    &28.6&11.53&22.75&31.08&28.52&53.17& 29.28& 11.22& 29.48 & 20.35    \\ \bottomrule   
\end{tabular}}
\vspace{0.2cm}
\caption{WER Results on different test sets using two different SSL backbone encoders; Data2Vec Base and Wav2Vec2 XLSR. The English fine-tuning is performed on the GigaSpeech ``XS" subset and the Danish one on $50$ hours of the NST dataset.}
\vspace{-0.8cm}
\label{tab:results}
\end{table*}

\section{Experiments and Results} \label{sec:exps}
This section outlines first the experimental details from dataset choices to used hyperparameters and configurations. It, then, presents and describes the obtained results. 
\subsection{Datasets} \label{sec:datasets}
The selected downstream sets are of reduced sizes as this work explores fine-tuning options in low-resource scenarios. We will evaluate our methods in two languages, English and Danish. For the English sets, GigaSpeech \cite{chen21o_interspeech} XS subset (10 hours) will be used for training instead of LibriSpeech \cite{librispeech} as the latter is in the pretraining sets, prohibiting proper forgetting considerations. The testing sets include the GigaSpeech test set, LibriSpeech test splits (test-clean and test-other), two datasets of Scottish and Welsh English accents \cite{demirsahin-etal-2020-open} and CommonVoice 14.0 English \cite{ardila-etal-2020-common} test set. The last three sets can be seen as the OOD testing samples as they present different accents and noise conditions. One may consider that OOD testing samples only underline a poor alignment between training and testing data. We argue that, in low-resource scenarios, with generally poor speaker and noise diversity, OOD use cases are almost inevitable, justifying our special interest in OOD performance.

For Danish, we use for training the NST Danish ASR Database \footnote{nb.no/sprakbanken/en/resource-catalogue/oai-nb-no-sbr-55/}. It consists of read speech samples recorded in very similar conditions,  thus enabling large possibilities for out-of-domain testing. $50$ hours of the dataset, from multiple speakers, are randomly selected for training, $5$ for validation and $10$ for in-domain testing. The exact splits are released within the code repository. The CommonVoice 14.0 Danish validation and testing splits are concatenated and used for OOD testing.  

\subsection{Self-Supervised Models}
Two self-supervised models are considered, Data2Vec Base \cite{baevski2022data2vec} and XLSR-53  \cite{conneau21_interspeech}. They offer variability in network size (90M parameters for the former and 317M for the latter), pretraining dataset diversity and size, and finally training loss and methods. Data2Vec Base is only trained on the LibriSpeech training splits, grouping $960$ hours of English read speech, while XLSR-53 is trained on a total of 56k hours of speech data covering 53 languages. 

\subsection{Methods Parameters} \label{sec:details}
This section gives the training details for the different continual-learning-based fine-tuning methods proposed in this work. 
\noindent \textbf{Baselines.}  The only hyperparameter for the baselines concerns the length of the freezing phase in the ``two-phased" approach, we fix it to 3 epochs.

\noindent \textbf{Freezing Based.} We use the LoRaLib toolkit \cite{hu2022lora} to replace the feed-forward layers following the transformer with a LoRa layer with rank $r$ as described in Section \ref{sec:methods}. We chose $r=16$ as in previous works on PEFT \cite{peftser}. For adapter architectures, we follow previous works in ASR \cite{eeckt2023using}, with a bottleneck linear layer followed by an upsampling one. Finally, applying Elastic Weight Consolidation requires two choices. First, we fix the hyper-parameter controlling the distance to the original model loss (see Equation \ref{eq:ewc}) to $\lambda= 50$. The Fisher information values are estimated on the LibriSpeech 10h split \cite{kahn2020libri}.

\noindent \textbf{Replay Based.} During the replay-based experiments, the pretraining tasks, masked latent prediction for Data2Vec, and contrastive predictive coding for XLSR are performed along with the ASR downstream one, as described in Section \ref{sec:methods}. During the first epoch of fine-tuning, no replay is done as it has been shown to lead to more stable fine-tunings. In the next epochs, fixing the replay probability to $p_R = 0.25$ led to the best results. The hyper-parameters of the replay task, mainly controlling the mask creation, are kept similar to the default ones used for the pretraining. Finally, replay requires the choice of a replay dataset. In the following, we will call ``auto-replay" experiments where the fine-tuning dataset is also used for the replay episodes. In a second experiment, either for English or Danish fine-tunings, replay batches will be sampled from LibriSpeech \textit{train} splits, as they are included in the training sets of Data2vec Base and XLSR-53. We call this experiment ``LS-replay". 

 \begin{figure*}[ht!]
  \centering
  \includegraphics[width=0.65\linewidth]{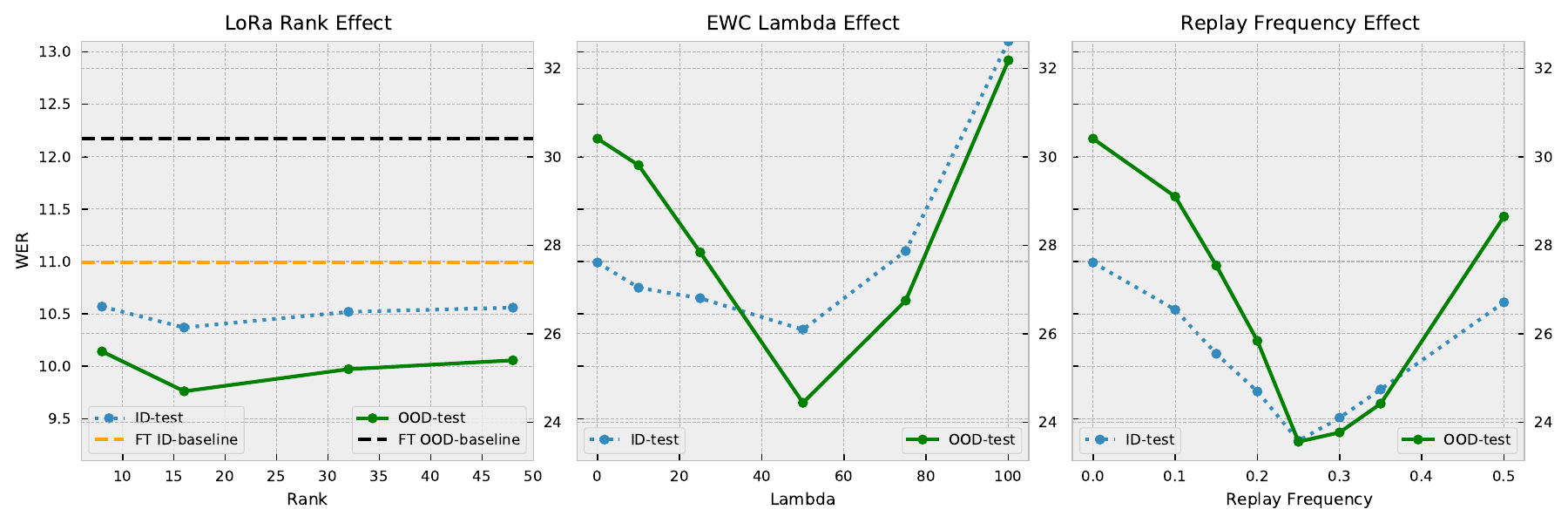}
  \caption{Effect of different hyper-parameters on the final performance on Danish in-domain (ID, left y-axis) and out-of-domain (OOD, right y-axis) test sets, for three different techniques (LoRa, EWC, and LS-Replay), with XLSR backbone. While LoRa seems quite robust to changes in the main hyperparameter, always remaining under the baseline, other approaches require careful tuning. In the second and third plots, the fine-tuning baseline is shown for $x=0$, while it is shown with horizontal dashed lines for the LoRa plot.   } 
  \label{fig:hp}

\end{figure*}
\subsection{Speech Recognition Settings} 
Two fully connected layers, with a hidden size of $1,024$ map each frame vector to one of the considered characters. The whole model is fine-tuned using the Connectionist Temporal Classification (CTC) loss as the downstream loss. During inference, greedy decoding is applied to the CTC probability outputs without any language-model-based re-scoring following the SpeechBrain recipe \cite{speechbrain}. More details are available in the accompanying GitHub repository.

 \begin{figure}[ht!]
  \centering
  \includegraphics[width=0.85\linewidth]{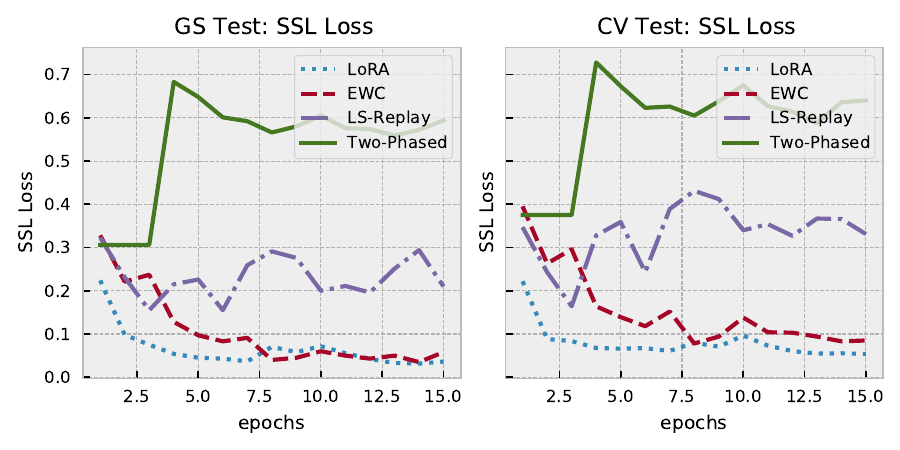}
  \caption{Evolution of the self-supervision task loss for 4 considered techniques on two English test sets with Data2Vec backbone. The best-performing approaches on the ASR task are the ones best-performing at the SSL task after the fine-tuning.} 
  \label{fig:sslloss}
\vspace{-0.7cm}
\end{figure}

\subsection{Results} \label{sec:results}

Table \ref{tab:results} shows the Word Error Rates (WER) obtained in the English and Danish experiments, with Data2Vec for the upper part and XLSR in the bottom part, as the backbone self-supervised representation model. Results for the English training, \textit{i.e.} the training performed on GigaSpeech XS, are shown on the five test sets described in Section \ref{sec:datasets}, while results for the Danish one are shown on two test sets. Every number shown is the mean of three runs with three different random seeds.

\noindent \textbf{Baselines.} As expected, the frozen model leads to poor performance. It is the worst-performing approach for both languages and with both Data2Vec or XLSR. Even worse, the model is not able to fit with XLSR with frozen features. The two classic baselines, freezing the convolutional front-end and the two-phased training, seem to perform better than the full fine-tuning baseline, especially for out-of-domain samples. We can see for instances in Table \ref{tab:results} an absolute gain of $3.5\%$ WER  and $2.3\%$ on CommonVoice English and Danish with the ``fixed CNN" approach compared to the full fine-tuning approach. 

\noindent \textbf{Freezing-based.} When considering the lower parts of the two blocks, presenting the alternative fine-tuning approaches results, we can see that, except for the failing ``adapters" approach, all the methods lead to better performances, both for in-domain and out-of-domain testing cases. This is visible from the numbers in bold in the table, as for every test set, the best performance is systematically obtained from one of the proposed alternatives. For instance, Low-Rank fine-tuning, while also being more efficient during training, achieves a mean error rate $7.0\%$ lower with Data2Vec and even $14.2\%$ lower with Wav2Vec2 XLSR. In a few settings, we failed to achieve reasonable performance with adapters and leave exploration of better adapter options for future works.

\noindent \textbf{Replay-based.} The replay-based approaches show two rows, ``LS-replay" and ``Auto-replay", as described in Section \ref{sec:details}, depending on the replay dataset, either LibriSpeech (LS) or the fine-tuning set. In all our settings, with both SSL backbones and on both target languages, replaying LibriSpeech samples, instead of target ones, leads to lower WERs. ``LS-replay" is even the best overall performing approach in two cases, Scottish and Welsh accented samples with Data2Vec Base and all Danish test sets with XLSR. The second case is surprising as the downstream and replay data are in different languages.

\noindent \textbf{OOD Generalization.} The two CommonVoice (CV) columns allow us to have a proper look at out-of-domain generalization. CV is a crowd-collected dataset showing various accents and recording conditions. This explains in part the high WER values in these columns. Compared to the full fine-tuning baseline, different freezing or replay-based approaches, allow a relative gain in performance that can reach $9.4\%$ and $14.8\%$ with Data2Vec Base, respectively for English and Danish. Relative gains even reach $15.7\%$ and  $22.5\%$ for Wav2Vec2 XLSR.

\section{Analysis and Discussion} \label{sec:analysis}
This section examines, first, the link between gains in performance and pretraining forgetting. Second, it discusses the sensitivity to hyperparameters of the proposed approaches.
\subsection{Probing the Forgetting}
To diagnose the link between the performance gains and non-forgetting of the self-supervision part, a checkpoint of the model at every epoch of fine-tuning is saved for probing. We compute, on a selected test set, the Data2Vec self-supervised task loss obtained with the fine-tuned model, at every saved model checkpoint. Data2Vec is trained to predict the latent representations of masked speech parts. During the probing, we mask parts of the input after the convolutional front-end, and see whether the model recovers the latent representations of a non-masked audio.  The self-supervision loss is computed on two English test sets, the in-domain test split of GigaSpeech and the OOD CV. 

We choose to show the three best-performing approaches, LoRa, EWC, and LS-Replay, along with the best baseline, the two-phased fine-tuning in Figure \ref{fig:sslloss}. The two-phased baseline shows an outlying behavior with a loss value twice as high as the ``LS-replay" and $6$ to $7$ times as high as for LoRa and EWC, after the frozen start of 3 epochs. This probing experiment seems to confirm the starting postulate with less forgetting being correlated with higher performance.

\subsection{Sensitivty to Hyperparameters} \label{sec:caveats}
As discussed in Sections \ref{sec:methods} and \ref{sec:details}, the presented fine-tuning approaches introduce various hyperparameters and choices. We highlight their influence by reporting the results of a group of experiments related to the tuning of these hyperparameters for the best-performing set of techniques. We consider for low-rank fine-tuning, EWC and ``LS-replay" the most impacting hyperparameter: respectively, the rank $r$ of the LoRa layers, the $\lambda$ parameter controlling the weight of the distance penalization in EWC and the frequency $p_R$ of replay episodes during fine-tuning. We show the results with different values of these hyperparameters on the in-domain (NST) and out-of-domain (CommonVoice) testing samples for the Danish training performed with XLSR-53 as the SSL backbone in Figure \ref{fig:hp}. 

For LoRA, the final performance is not severely impacted by reasonable changes in the main hyperparameter, the rank of the Lora layer. However, this is not the case for the two other techniques, as shown clearly with the ``V"  shapes of the plots in the second and third columns of Figure \ref{fig:hp}. With inappropriate values, word error rates are higher than the full FT baseline.

\section{Conclusion}
This work tests continual-learning-inspired fine-tuning approaches for self-supervision-based speech recognition. Results show that LoRA fine-tuning, EWC, and replay allow substantial gains compared to the full fine-tuning baseline, reaching $14.8\%$ and $22.5\%$ on Danish ASR, with two different SSL encoders. These gains are correlated with less forgetting, \textit{i.e.}, better performance on the pretraining task after fine-tuning.

\bibliographystyle{IEEEtran}
\bibliography{mybib}

\end{document}